\documentclass{nadastro}
\usepackage{times}
\usepackage{overcite}
\newcommand{\SNLMC}{\mbox{SN$\,$1987A}$\,$}
\newcommand{\xmn}[2]{\mbox{$#1\!\times\! 10^{#2}\,$}}
\newcommand{\avr}[1]{\mbox{$\langle #1\rangle$}}
\newcommand{\gsim}{\:\raisebox{.25ex}{$>$}\hspace*{-.75em}
      \raisebox{-.93ex}{$\sim$}\:} 
\newcommand{\lsim}{\:\raisebox{.25ex}{$<$}\hspace*{-.75em}
      \raisebox{-.93ex}{$\sim$}\:}
\newcommand{\isn}[2]{\mbox{$^{#2}${#1}}}

\newcommand{\MS}{\mbox{$M_{\textstyle\odot}\,$}}
\newcommand{\NMS}[1]{\mbox{$#1\,M_{\textstyle\odot}\,$}}
\newcommand{\RS}{\mbox{$R_{\textstyle\odot}\,$}}
\newcommand{\NRS}[1]{\mbox{$#1\,R_{\textstyle\odot}\,$}}
\begin{document}

\markboth{D.~K.~Nadyozhin \& V.~S.~Imshennik} 
 {Physics of Supernovae}

%
\catchline{}{}{}{}{}
%
\title{PHYSICS OF SUPERNOVAE\footnote{Invited lecture at the
19th European Cosmic Ray Symposium, 
August 30 -- September 3, 2004, Florence, Italy}}

\author{\footnotesize D.~K.~NADYOZHIN and V.~S.~IMSHENNIK}
\address{A.~I.~Alikhanov Institute for Theoretical and Experimental 
         Physics \\
         B. Cheremushkinskaya St.\ 25, RU-117218, Moscow, Russia}

\maketitle

\pub{ }{ }

\begin{abstract}
 The origin of cosmic rays (CR) is supposed to be closely connected
 with supernovae (SNe) which create the conditions favorable for 
 various mechanisms of the CR acceleration to operate effectively.
 First, modern ideas about the physics of the SN explosion 
 are briefly discussed: the explosive
 thermonuclear burning in degenerate white dwarfs resulting in 
 Type Ia SNe and the gravitational collapse of 
 stellar cores giving rise to other types of SNe (Ib, Ic, IIL, IIP).
 Next, we survey some global properties of the SNe of 
 different types: the total explosion energy distribution of 
 various components (kinetic energy of the hydrodynamic flow, 
 electromagnetic radiation, temporal behavior of the neutrino 
 emission and individual energies of different neutrino flavors).
 Then, we discuss in the possibility of direct hydrodynamic 
 acceleration by the shock wave breakout and 
 the properties of the SN shocks in the circumstellar medium. 
 Then the properties of the neutrino radiation from the
 core-collapse SNe and a possibility to incorporate both 
 the LSD Mont Blanc neutrino event and that recorded by 
 the K~II and IMB detectors into a single
 scenario are described in detail. 
 Finally, the issues of the neutrino nucleosynthesis
 and of the connection 
 between supernova and gamma-ray bursts are discussed.

\end{abstract}

\section{Basic Properties of Supernovae}

 Physically, there are two fundamental types of supernovae
 (SNe): the thermonuclear SNe and the core-collapse ones, 
 represented by Type Ia SNe (SN~Ia) and by Type II, Ib, 
 and Ic SNe, respectively. 
 The SN~Ia show no hydrogen in their spectra and constitute
 quite a homogeneous sample of objects.
 The core-collapse SNe are subdivided into several types
 depending on the amount of hydrogen hanging around the stellar
 core just before it begins to collapse. Type Ib and Ic SNe 
 have virtually no hydrogen left. The Ic SNe seem to have 
 lost a fair amount of their helium as well. 
 
 Type II SNe are represented by the subtypes IIP 
 (plateau shaped light curves, \NMS{\sim 10} of hydrogen in 
  their envelopes), IIL (linearly decaying light curves, 
 \NMS{\lsim 1} of hydrogen), and IIn (with some hydrogen 
 in an extended envelope formed by dense stellar wind). 
 
\subsection{Thermonuclear supernovae}
 
 The SN~Ia light curves are powered by the 
 $\isn{Ni}{56}\rightarrow\isn{Co}{56}\rightarrow\isn{Fe}{56}$
 beta-decay on average of $\NMS{\approx0.6}$ of \isn{Ni}{56}
 synthesized as a result of explosive carbon-oxygen (CO) burning 
 in a degenerate {\em Chandrasekhar mass\/} $(M\NMS{\approx1.4})$
 white dwarf. The total energy of the electromagnetic emission 
 is of \xmn{\approx 6}{49}erg, most of this energy being radiated
 in optical and infrared wavelengths and only a fraction being 
 carried away by X-rays and gamma-photons that managed to escape 
 the scattering off by the expanding envelope.
 The explosion energy $E_{\mathrm{exp}}\approx 10^{51}\,$erg
 comes from the difference in nuclear binding energies of 
 the initial carbon-oxygen mixture and the final products of 
 thermonuclear burning (mainly \isn{Ni}{56} -- the most tightly 
 bounded nucleus among all those consisting of equal numbers 
 of neutrons and protons). Finally, almost all $E_{\mathrm{exp}}$ 
 turns out to be converted into the kinetic energy of expanding
 matter. The white dwarf proves to be totally disrupted by 
 the explosion, no stellar remnant being left.
 The mean velocity of the expanding debris is estimated to be 
 \avr{u}$=\sqrt{2E_{\mathrm{exp}}/M}\approx 8,000\,$km/s.
 
 Although observationally and theoretically the above concept 
 is considered to be a well-founded conjecture, there remains 
 a big unsolved problem relating to the mode of thermonuclear
 CO-burning. The most important issue is an interplay between
 the deflagration and detonation regimes of burning which is
 controlled by different instabilities of turbulent
 thermonuclear flame propagating in degenerate matter and by 
 the behavior of the white dwarf as a whole in response to 
 the onset of the burning (pre-expansion\cite{Kho91,WoWe94},
 radial pulsations\cite{IIC74,DuImBl01}). Several astrophysical
 groups are currently engaged in an extensive investigation
 of the thermonuclear burning in Type Ia supernovae 
 (see Refs.~\refcite{WWK04,RoHi04,GKO04} and references therein).
 This complicated problem requires a sophisticated approach
 based on three-dimensional modeling of the CO ignition and
 propagation of the thermonuclear flame that is fraught with 
 specific unsteadinesses such as convective, Rayleigh-Taylor,
 Landau--Darries instabilities.
 
 Such an investigation is of fundamental importance for
 the accurate calibration of SN Ia as the cosmological 
 standard candles (one needs to predict the SN Ia peak
 luminosity at least with a 10\% precision!). 
 Also for detailed comparison with observations 
 and for stellar nucleosynthesis, a precise 
 determination of different isotopic yields 
 (apart from dominating \isn{Ni}{56}) is of vital 
 importance.
 
\subsection{Core-collapse supernovae}

 Type II SNe light curves are powered first by 
 the shock heating, then by recombination of hydrogen, 
 and finally (at their {\em tail phase}) by the
 $\isn{Co}{56}\rightarrow\isn{Fe}{56}$ decay of 
 $\approx$\NMS{(0.02-0.2)} of \isn{Co}{56} 
 (initially \isn{Ni}{56}). The total energy of 
 the electromagnetic emission is of $\approx 10^{49}\,$erg.
 
 The explosion energy of the core-collapse SNe is 
 typically of \xmn{(0.5-2)}{51}$\,$erg. It comes 
 from the shock wave that is launched somewhere at 
 the boundary between the ``iron'' core of a mass
 $M_{\mathrm{Fe}}=\NMS{(1.2-2)}$, collapsing into 
 a neutron star (NS), and the outer envelope to be thrown out.
 The mean velocity of the expansion is of 
 $3,000-5,000\,$km/s depending on the mass of the envelope
 expelled. 
 
 The mechanism of the core-collapse SNe is not yet 
 understood in every detail. The most distinctive feature
 of these SNe is an enormous energy of 
 \xmn{(3-5)}{53}erg$\, =(10-15)$\%\,$M_{\mathrm{Fe}}c^2$ radiated
 in form of neutrinos and antineutrinos of all the flavors
 $({\mathrm e},\mu,\tau)$. 
 One would think that it should not be a problem to extract less 
 than 1\% from the energy of a powerful neutrino flux to ensure
 the expulsion of the SNe envelope.
 However, an extensive hydrodynamical modeling during the last 
 thirty years has demonstrated that in case of spherical symmetry
 it is very hard (if not impossible) to simulate the explosion.
 Basing on this research, an empirical theorem can be formulated
 telling that spherically-symmetrical models
 do not result in expulsion of an envelope; the SN outburst does 
 not occur: the envelope falls back on the collapsed core. 
 Hence, one has to go to two- and, perhaps, three-dimensional models 
 to convert the stalled accreting shock into an outgoing blast wave
 (for the last review see Ref.~\refcite{BWOL04} and references therein).
 One has, nevertheless, to keep in mind that yet undiscovered
 elementary particles may be involved in the core-collapse SNe
 (e.g., axion-like particles\cite{Berezh03}).
 
 There are three reasons of spherical symmetry breakdown which 
 currently are under a close investigation:
 \begin{itemlist}
 \item  {\sl Large-scale neutrino-driven convection\/}.\\
        The accreting shock can obtain an additional energy 
        necessary for successful explosion from fast  
        (possibly jet-like) subsonic streams of neutrino-heated matter 
        circulating under and over the neutrinosphere\cite{Eps79,LiBuCo80,LaMa81,Imsh02}.
 \item {\sl Interaction between rotation and magnetic field\/}.\\
       Hydrodynamical heterogeneity of the collapse (central dense
       layers of stellar core contract increasingly faster than
       outer ones) results in a strong differential rotation that
       leads to an amplification of toroidal magnetic field.
       Under favorable conditions, an excessive magneto-hydrodynamical 
       pressure could facilitate the expulsion of the supernova 
       envelope\cite{Bis1970,MBA2003}.
 \item {\sl Rotational fragmentation followed by a NS explosion\/}.\\
     Massive fast-rotating collapsed core undergoes  rotational
     fragmentation resulting in formation of a close neutron-star binary 
     that evolves being driven by the emission of gravitational waves and
     mass-exchange and terminates with a low-mass $(M\approx\NMS{0.1})$
     neutron-star explosion\cite{Imsh92,ImPop94}. 
 \end{itemlist}

 \begin{figure}
 \centerline{
\epsfig{file=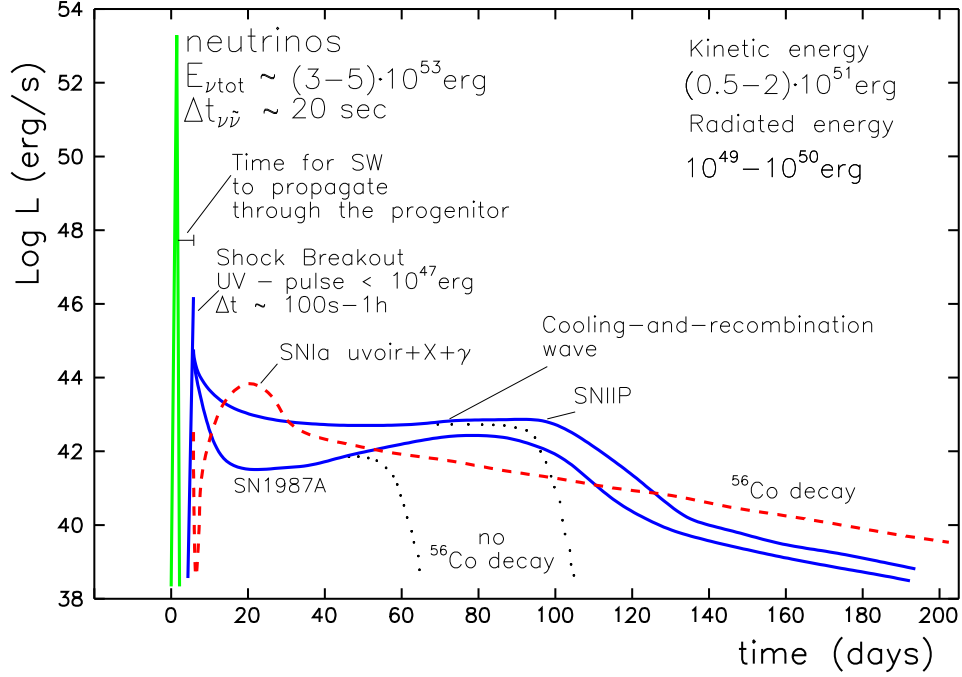,width=\textwidth}
 } 
 \vspace*{8pt} 
 \caption{A schematic illustration of the light curves and other
          supernova properties.}
 \label{lcurvs}
 \end{figure}
 
  Figure~\ref{lcurvs} shows a general view of electromagnetic and neutrino 
  luminosity of SNe.  The SNIa bolometric light curve is given by 
  a dashed line that includes all electromagnetic spectrum 
  (uvoir $+ X + \gamma$: ultraviolet, optical, infrared, $X$-rays, 
  and gamma radiation). In $\sim 40\,$days after explosion, the
  light curve strictly follows the \isn{Co}{56} decay
  half-life of 77.1 days (111.3 days for exponential decay time). 
  There is shown also the typical SNIIP light curve with 
  a $\sim$100-day period of nearly constant luminosity (plateau) 
  stabilized by a cooling-and-recombination wave\cite{GIN1971,Nad94}.
  If there were no \isn{Co}{56} in the supernova envelopes 
  the light curves would be of a shorter duration (dotted curves).
  In the case of \SNLMC in the Large Magellanic Cloud, 
  about \NMS{0.075} of \isn{Ni}{56} was synthesized.

\begin{figure}
 \centerline{
 \epsfig{file=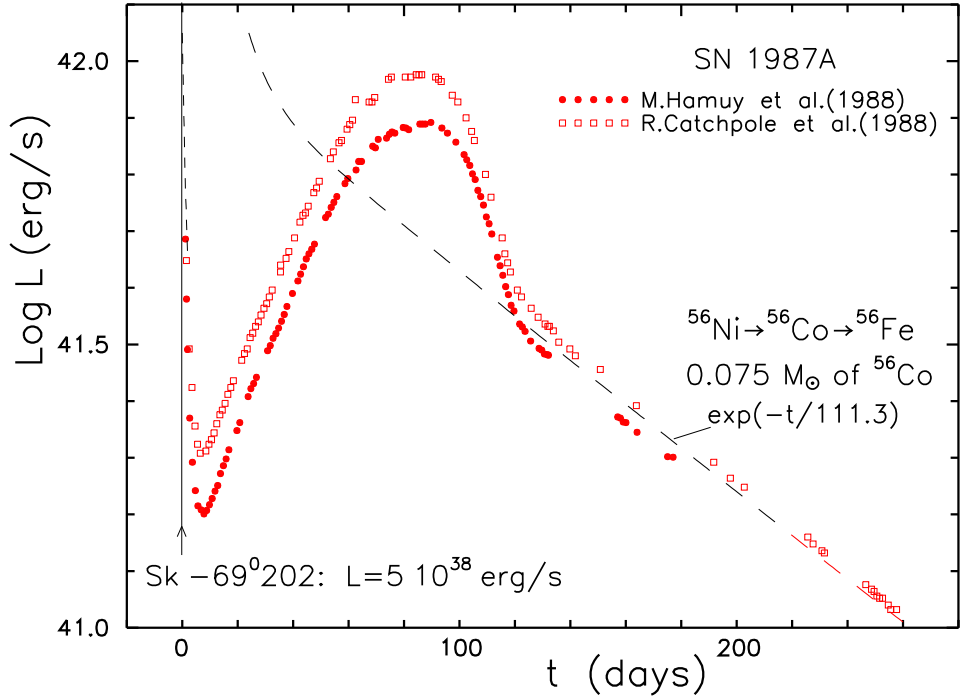,width=\textwidth} 
 } 
 \vspace*{8pt} 
 \caption{The bolometric light curve of \protect\SNLMC exploded 
  on February 23, 1987. Time is measured from the moment
  of the shock wave breakout.
  At $t\approx 90\,$days, the \SNLMC luminosity attains 
  a maximum of $\approx 10^{42}$erg/s that by a factor 
  of 2,000 exceeds the luminosity of the progenitor, 
  blue supergiant Sk$-69^o202$. The shock wave breakout ``tail''
  is shown by a nearly vertical dashed line at $t\approx 0$.
  (Adapted from Ref.~\protect\refcite{Nad94}).}
  \label{lcSNA}       
 \end{figure}

    Figure~\ref{lcSNA} shows the \SNLMC light curve on a large scale
    as observed by two group of astronomers in Chili\cite{Ham88}
    (black dots) and South Africa\cite{Catch88} (open squares).
    A 20\% discrepancy between these two sets of observations 
    is a systematic uncertainty connected with reconstruction 
    of the bolometric luminosity from luminosities observed 
    in different spectral bands.
    The coincidence of the \SNLMC bolometric light curve 
    with the Co-decay law at $t>140\,$days gave the first 
    direct proof that supernova ejecta are actually 
    enriched with theoretically predicted \isn{Co}{56}.
    In a month, this finding was confirmed by 
    the detection of X-rays by space-based detectors 
    {\sl Kvant\/}\cite{Sunya87} and {\sl Ginga\/}\cite{Ginga87}
    and somewhat later by direct measurement of gamma-lines
    from Co-decay\cite{Smax88}. Detailed report on the \SNLMC
    event can be found in special 
    reviews\cite{ImNad89,Arnet89,Nad92}. For recent thorough study
    of the \SNLMC light curve on the basis of radiation hydrodynamics
    with nonthermal ionization from the \isn{Ni}{56} and \isn{Co}{56}
    decays included see Ref.~\refcite{Utr04}.
      
\section{Shock Wave Breakout}

\begin{figure}
 \centerline{
 \epsfig{file=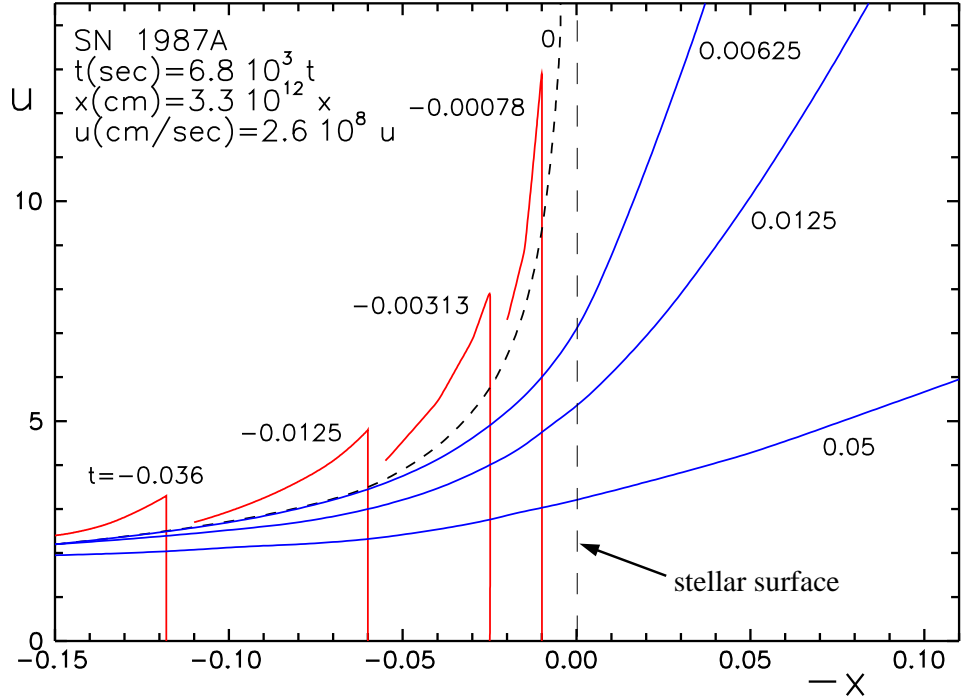,width=\textwidth}
 } 
 \vspace*{8pt} 
 \caption{Velocity $u$ versus distance $x$ from stellar surface 
          at different points of time $t$ during the shock wave
          breakout. It is assumed that $t=0$ when the shock reaches
          the surface (short-dashed curve).
          The equations one must use to convert 
          $u$, $x$, and $t$ to dimensional units 
          (cm, sec) are shown for the case of \protect\SNLMC.
          (Adapted from Ref.~\protect\refcite{Nad94}).}
 \label{swbout}
 \end{figure}

 The onset of supernova outburst occurs at the very time
 when the outgoing shock wave (SW) reaches the stellar
 surface. Such a breakout results in a short pulse of ultraviolet
 and soft X-ray radiation of total energy up to $10^{47}\,$erg
 and with characteristic duration of 100s--1h depending on 
 the presupernova radius (Fig.~\ref{lcurvs}). The ``tail" of this pulse 
 was actually observed in the case of \SNLMC (Fig.~\ref{lcSNA}).
 The shock wave propagates through stellar interior outward 
 in the direction of strongly decreasing density.
 Consequently, the shock energy turns out to be accumulated in
 a progressively decreasing amount of matter.
 As a result, the SW considerably accelerates as illustrated 
 in Fig.~\ref{swbout}. 
 Such a cumulative regime is described by well-known similarity
 solution of hydrodynamic equations that exists since the density
 $\rho$ typically is a power function of the distance $x$ 
 to stellar surface $\rho\sim x^n$ ($n\approx 3$).
 Formally, the velocity tends to infinity at stellar surface
 (short-dashed curve for $t=0$). In reality, the SW cumulation 
 is limited by a finite width of the SW front which optical
 thickness $\Delta\tau$ for the radiation dominated SW can be 
 estimated from a simple relation: $\Delta\tau\approx c/D$,
 with $D$ and $c$ being the SW velocity and the speed of light,
 respectively. The distance $x_{\mathrm{cut}}$ at which one 
 has to cut the similarity solution off corresponds to
 the SW position when its optical depth becomes just equal to
 $\Delta\tau$ (see Ref.~\refcite{ImNad88}).
 With $x_{\mathrm{cut}}$ known one can estimate
 the maximum velocity at the SW front emerging from under the stellar
 surface. During further expansion (curves for $t>0$ in Fig.~\ref{swbout}),
 matter undergoes additional acceleration in a rarefaction wave
 that converts almost all thermal energy into kinetic energy 
 of radial expansion increasing the latter 
 by a factor of $\approx 2.6$ (Ref.~\refcite{LiNa90}).
 
  For \SNLMC, $x_{\mathrm{cut}}$ is estimated to be $\approx 0.02R_0$ 
  ($R_0\approx\NRS{47}$ is the presupernova radius). The resulting 
  final maximum velocity $u_{\mathrm{max}}$ relating to
  kinetic energy per nucleon 
  $\varepsilon_{\mathrm{max}}=\frac{1}{2}m_{\mathrm u}u_{\mathrm{max}}^2$
  ($m_{\mathrm u}$ is the atomic mass unit) 
  and the mass $\Delta M_{u\mathrm{max}}$ accelerated to the maxim 
  velocity $u_{\mathrm{max}}$ turn out to be \cite{Nad94}:
  $$ u_{\mathrm{max}}\approx 40,000\,\mbox{km/s}\, ,\quad 
     \varepsilon_{\mathrm{max}}\approx 8.3\,\mbox{MeV/nucleon},\quad
     \Delta M_{u\mathrm{max}}\approx\xmn{2}{-6}\MS.$$
     
   The SW breakout was expected\cite{ColJh60} to be an efficient
   mechanism for acceleration of cosmic rays. For \SNLMC
   this mechanism, however, does not look effective enough.
   Since the maximum kinetic energy changes with $R_0$ as 
   $R_0^{-0.65}$, one can think of the SN explosions associated 
   with presupernovae of smaller radii.
   The SN~Ib and SN~Ic can have as small $R_0$ as a few \RS and
   $\varepsilon_{\mathrm{max}}$ may reach about 100\,MeV/nucleon 
   for these SNe. The explosion of a white dwarf of typical radius
   $(5,000-10,000)\,$km would be just the right event to accelerate
   a good portion of matter to relativistic energies 
   $(\varepsilon_{\mathrm{max}}\gsim 1\,$GeV/nucleon). 
   However, the regime of the SW cumulation does not occur in
   the case of SNe Ia that come from white dwarf progenitors.
   The thermonuclear burning begins there in a deflagration regime
   causing the star to expand subsonically. Even though the SW
   does appear this happens at the very end of the explosion
   under the conditions unfavorable for the SW cumulation.
   
   Although, to all appearance, the direct hydrodynamical 
   acceleration of CR in supernovae turns out to be inefficient
   the SW breakout could provide fast moving particles for
   further acceleration by other mechanisms 
   (e.g., by circum-stellar and interstellar shock waves).
   
\section{Shock Waves in Circum-Stellar Medium}

 \begin{figure}
 \centerline{
\epsfig{file=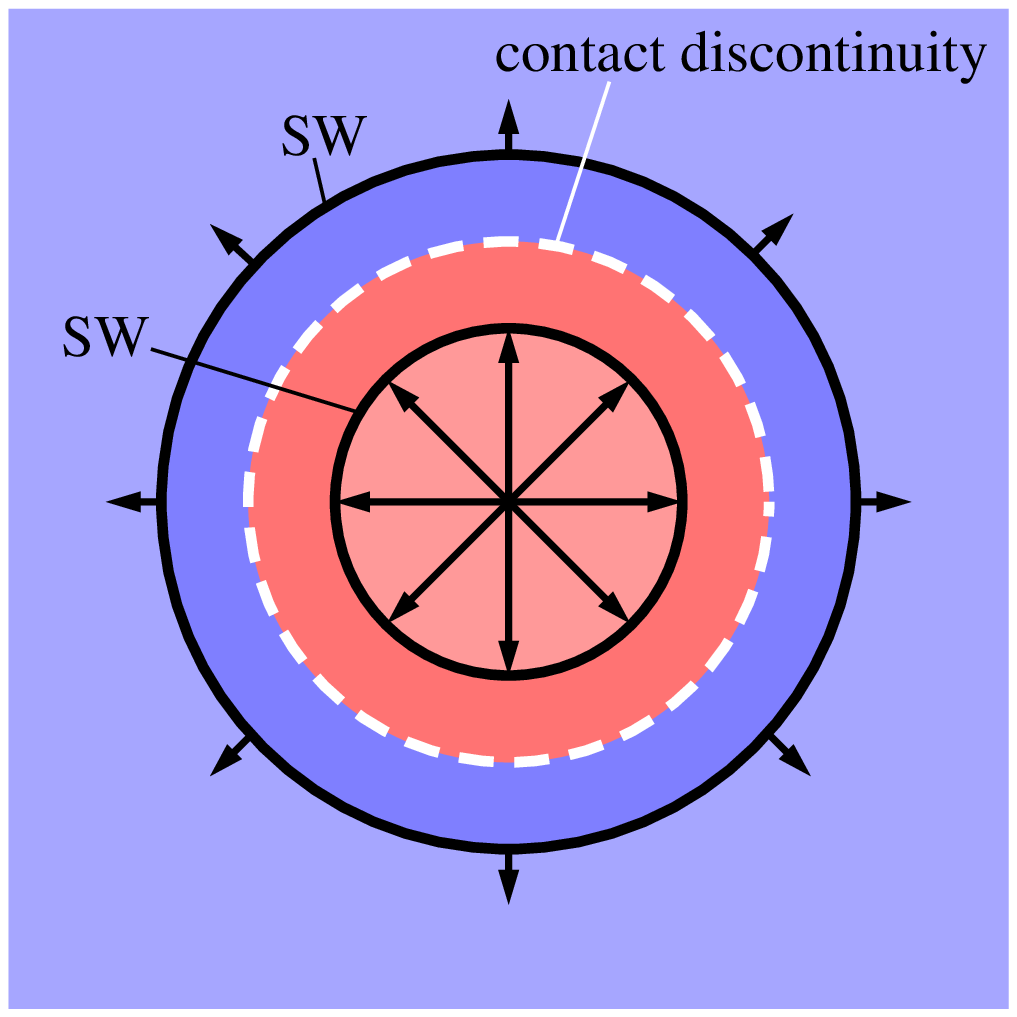,width=0.45\textwidth}\hfill
\epsfig{file=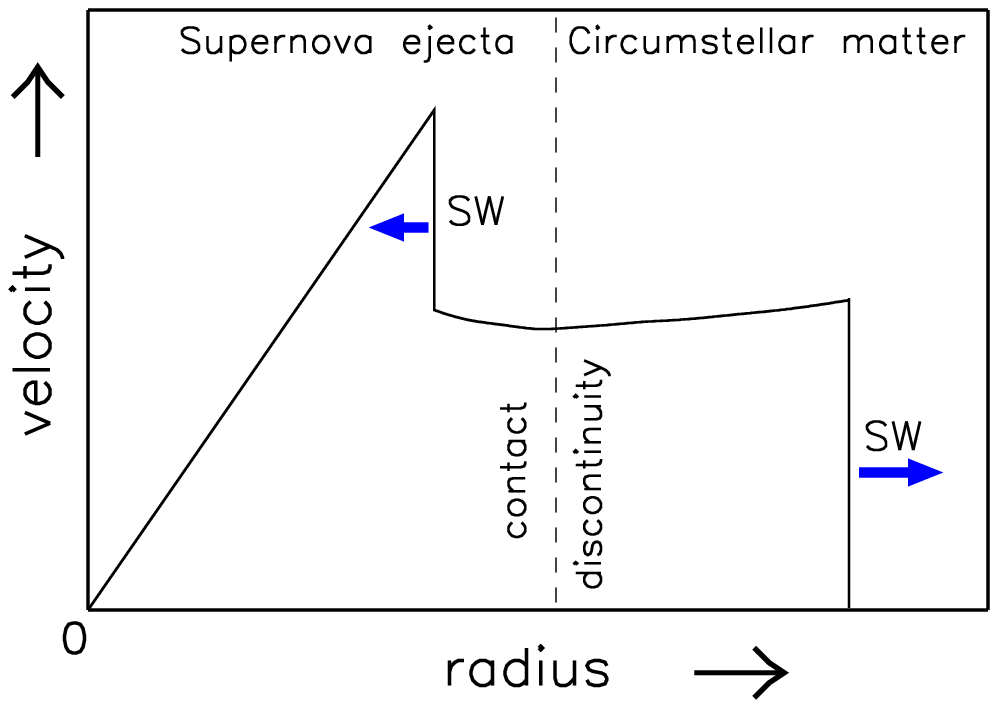,width=0.54\textwidth}
 } 
 \vspace*{8pt} 
 \caption{A schematic illustration of the interaction between 
  the SN ejecta and ambient medium: general hydrodynamic layout
  (left panel) and velocity versus radius in arbitrary units
  (right panel).}
  \label{circumst}
 \end{figure}
 \noindent
 In a few days after explosion, the SN envelope turns into a state
 of supersonic inertial expansion with the velocity depending on
 radius by the simplest way: $u=r/t$ where the time $t$
 is measured from the beginning of the explosion. Simultaneously
 the outer edge of the SN envelope begins to collide with 
 the circum-stellar matter. The interaction region is confined 
 by two shock waves shown in Fig.~\ref{circumst}
 (left panel) by black circles.
 The SN ejecta are decelerated, heated, and compressed by 
 the internal (reverse) SW. The outer (forward) SW accelerates, 
 heats and compresses the ambient medium. 
 Thus, {\sl relative to matter\/} the reverse and forward shock 
 waves propagate in an inward and outward direction, 
 respectively. This is shown by arrows in Fig.~\ref{circumst}
 (right panel).
 The interface separating the shocked ejecta from shocked 
 interstellar matter called {\sl contact discontinuity\/} 
 is shown by a white dashed circle in 
 Fig.~\ref{circumst} (left panel). 
 
  The hydrodynamic theory of the SN ejecta -- ambient medium 
  interaction has been elaborated with the help of numerical 
  and semi-analytical methods 
  (see Ref.~\refcite{TMK99} and references therein).
  The initial stage of this {\sl ejecta-dominated\/} process 
  is controlled by a similarity solution\cite{Cheva82,Nadezh85} 
  that being combined with observations allows to describe
  detailed structure of young supernova remnants (SNRs).
  A few hundred years old remnants of galactic supernovae 
  such as Cas A, Kepler, Tycho are still in the ejecta-dominated
  stage. Observations of their X-ray and radio synchrotron 
  emission can give much information about the SN progenitor,
  supernova nucleosynthesis\cite{DTycho01}, ambient medium, 
  and mechanisms of the cosmic-ray acceleration.
\begin{figure}
 \centerline{
\epsfig{file=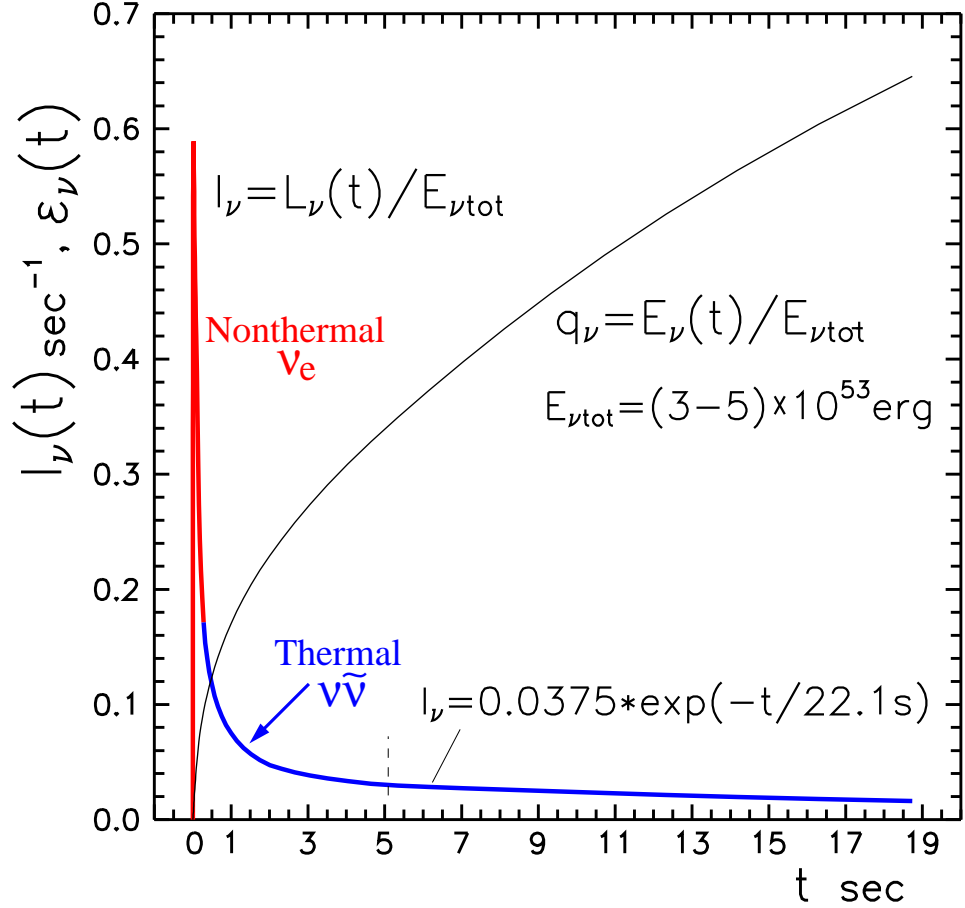,width=\textwidth}
 }
 \vspace*{8pt} 
 \caption{The normalized bolometric neutrino light curve.
     The time $t$ is measured from the beginning of the collapse.
    (Based on Ref.~\protect\refcite{Nadezh78}).}
 \label{lnubol}
 \end{figure}
    
\section{Core-Collapse Neutrinos}

 The collapse of stellar iron-cores into a neutron star (NS)
 is followed by a high-power pulse of neutrino emission.
 Figure~\ref{lnubol} shows the cumulative (bolometric)
 neutrino light curve that includes all the neutrino and 
 antineutrino flavors. This light curve was calculated 
 for a spherically symmetrical collapse of a \NMS{2} stellar
 core\cite{Nadezh78}. 
 According to detailed modeling during the past few decades 
 of the neutrino transport in collapsing stellar cores,
 this light curve consists of two parts. 
 
 The first part
 ($t\lsim 0.5\:$s) relates to the nonthermal neutrino emission
 that is dominated by the electron neutrinos $\nu_{\mathrm e}$
 produced by the non-equilibrium neutronization. 
 For $t\lsim 0.5\:$s, the core is still transparent to 
 $\nu_{\mathrm e}$ emitted owing to electron captures 
 by nuclei and free protons. 
 The mean individual $\nu_{\mathrm e}$
 energy $\varepsilon_{\nu\mathrm e}$ turns out to be 
 $15-20\:$MeV. The nonthermal neutrinos carry away only
 a small fraction ($q_\nu\lsim 10$\%, 
 Fig.~\ref{lnubol}) of the total available energy
 $E_{\nu\mathrm{tot}}=\xmn{(3-5)}{53}$erg. 
 
 About 90\% of $E_{\nu\mathrm{tot}}$
 is emitted in the regime of thermal emission when the central
 region of the core becomes opaque to all the neutrino flavors.
 The neutrinos are decoupled from stellar matter at a surface
 called {\sl neutrinosphere\/}. In general, the neutrinosphere
 radius is different for different neutrino flavors, at least
 one has to consider two neutrinospheres --- one for
 the electron neutrinos and antineutrinos and another for
 $\mu$- and $\tau$-neutrinos and antineutrinos.
 Numerical modeling\cite{Bruen85,Bruen87} shows that 
 to a first approximation one 
 can assume that $E_{\nu\mathrm{tot}}$ is equally distributed
 over the neutrino flavors:
 $$ E_{\nu\widetilde\nu\mathrm{e}}\approx 
    E_{\nu\widetilde\nu\mu}\approx E_{\nu\widetilde\nu\tau}
    \approx\frac{1}{3}E_{\nu\mathrm{tot}}.$$
 The neutrino spectra are reproduced by the Fermi--Dirac 
 distributions slightly depressed\cite{NadOtr80} at high energies
 $\varepsilon\gg kT_{\nu\mathrm{ph}}$ 
 ($T_{\nu\mathrm{ph}}$ is the effective temperature of 
 the neutrinosphere). The mean individual neutrino energies
 and related $T_{\nu\mathrm{ph}}$ are
 $$\avr{\varepsilon}_{\nu\widetilde\nu\mathrm{e}}
   \approx (10-12)\,\mathrm{MeV},\qquad 
   \avr{\varepsilon}_{\nu\widetilde\nu\mu}\approx
   \avr{\varepsilon}_{\nu\widetilde\nu\tau}\approx
   25\,\mathrm{MeV},$$
 $$T_{\nu\widetilde\nu\mathrm{e}\mathrm{ph}}\approx 4\,\mathrm{MeV},
   \qquad T_{\nu\widetilde\nu\mu\mathrm{ph}}\approx
   T_{\nu\widetilde\nu\tau\mathrm{ph}}\approx 8\,\mathrm{MeV}.$$
\begin{figure}
 \centerline{
\epsfig{file=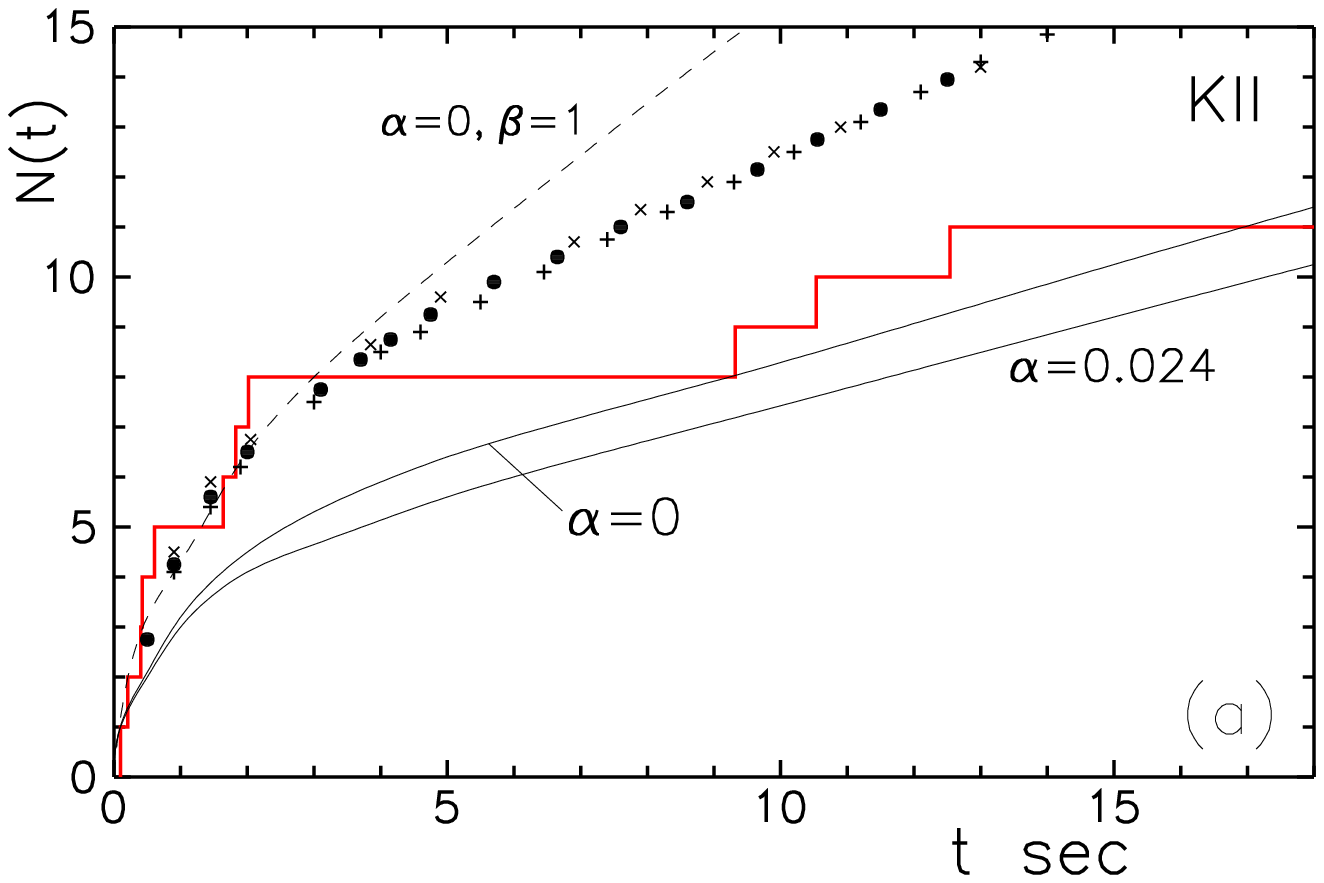,width=0.565\textwidth}\hfill
\epsfig{file=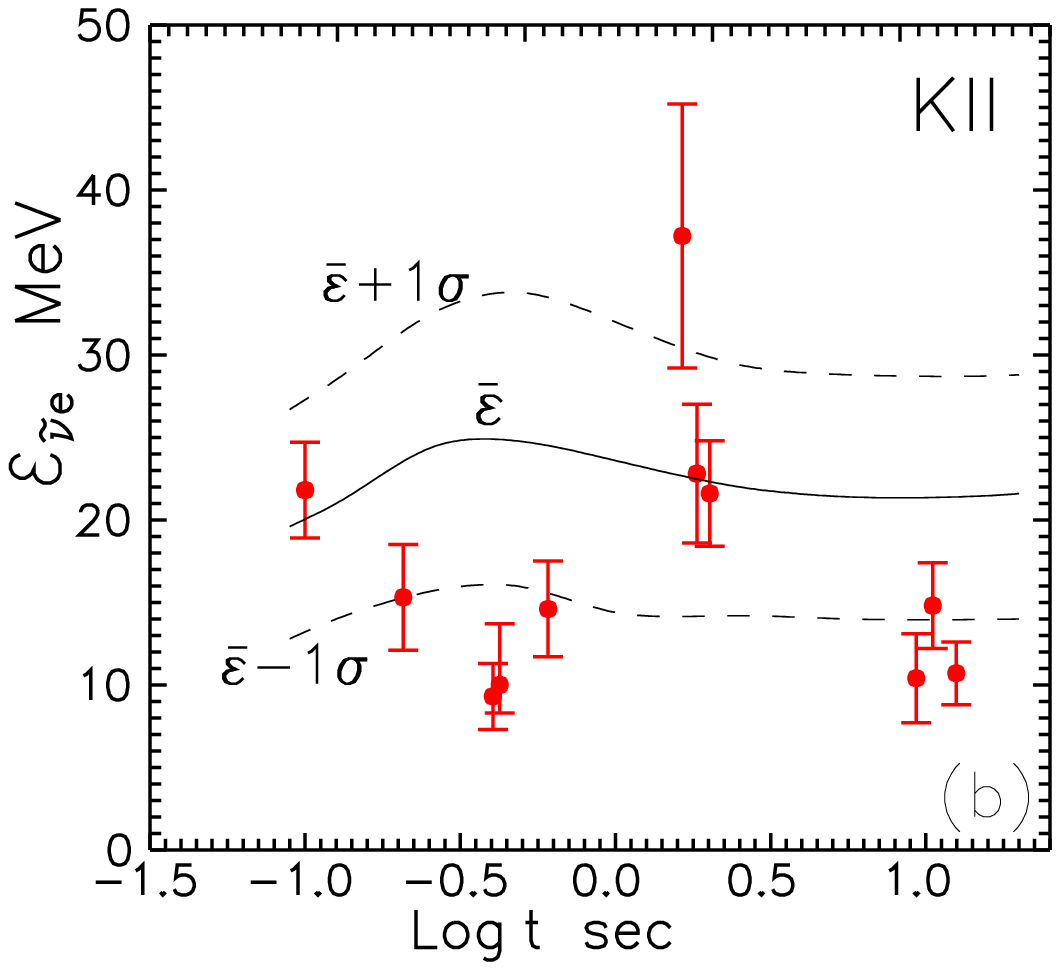,width=0.405\textwidth}}  
 \vspace*{8pt} 
 \caption{A comparison of the neutrino signal predicted for 
  \protect\SNLMC
  with the response of the KamokaNDE II detector\protect\cite{KIIdet}.
 ({\em a}) The number of counts versus time in the detector 
           (in total 11 events, step line). Different theoretical versions
           are also shown (see Ref.~\protect\refcite{ImNad89} for details).
 ({\em b}) The energy measured for every count in comparison with
           a theoretical prediction $\bar\varepsilon$ bounded by $\pm 1\sigma$
           uncertainty band.
           (From Ref.~\protect\refcite{ImNad89}).}
 \label{Kamioka}
 \end{figure}
  The characteristic time of the neutrino pulse turns out to be 
  of the order of 10--20$\,$s.
 
  These theoretically predicted properties seem to be
  in a fair agreement with the neutrino signal detected
  from \SNLMC by the KamiokaNDE II (K II)\cite{KIIdet}
  and Irvine-Michigan-Brookhaven (IMB)\cite{IMBdet} neutrino 
  detectors as it is shown in Fig.~\ref{Kamioka} for K II.
  The agreement occurs under the assumption that all the events
  in these water Cherenkov detectors were produced
  by the electron antineutrinos through the reaction
  \begin{equation}
  \widetilde\nu_{\mathrm e}+p\rightarrow n+e^+\, .
  \label{nupne}
  \end{equation}
  In contrast to the neutrino-electron scattering,
  the relativistic positrons from this reaction move virtually
  isotropically in different directions. So the direction
  of their Cherenkov radiation should not depend on where
  $\widetilde\nu_{\mathrm e}$ has come from.
  However, one can observe that the events with energies 
  $\varepsilon_{\widetilde\nu\mathrm{e}}\gsim 20\,$MeV 
  (four events in Fig.~\ref{Kamioka} and all eight events 
  recorded by the IMB detector) demonstrate 
  a statistically significant correlation with 
  the direction to the Large Magellanic Cloud. 
  Since the neutrino-electron scattering cross-section 
  is considerably lower than that of the reaction (\ref{nupne})
  it is impossible to explain this observation by addressing
  to the $\nu_{\mu\tau}$-electron scattering ---
  the required total energy of the neutrino pulse would exceed
  by an order of magnitude the energy available from 
  the collapse of stellar cores. The significance of this 
  problem (remaining unresolved up to now!) was first 
  recognized and analyzed in Refs.~\refcite{RR88,DZR89}.
  
  Another difficulty in the theoretical interpretation of
  the \SNLMC neutrino signal is the fact that there occurred 
  {\sl two\/} neutrino pulses. The first pulse, detected
  by the Liquid Scintillation Detector (LSD) 
  under Mont Blanc\cite{Dadyk87,Agli87},
  came 4.7 hours earlier than the second one recorded by
  the K~II and IMB detectors. Since 4.7$\,$h is a very long
  time in comparison with the duration of both the neutrino 
  pulses ($\sim 10\:$s), one has to think of a two-stage 
  collapse (see discussion in Ref.~\refcite{ImNad89}).
  In the next section, we describe a scenario that has
  been recently proposed\cite{ImRya04} to incorporate 
  both the neutrino pulses in a self-consistent two-stage
  hydrodynamical model of the gravitational collapse.
  
\section{Rotational Fragmentation---Neutron Star Explosion Scenario}

  The key point for this scenario is the presence of rotation
  in the stellar core that is about to collapse.
  The mechanism of the SN explosion proposed in Ref.~\refcite{Imsh92}
  is based on the rotational instability and develops through 
  the following stages.

  First, the rotational energy of the collapsing core 
  $E_\mathrm{rot}$ reaches the limit of  stability with respect to 
  fragmentation\cite{Tass78}: $E_\mathrm{rot}/|E_\mathrm{g}| >0.27$ 
  ($E_\mathrm{g}$ is the core gravitational energy).

 Then the core of mass $M_0$ fragments into a close binary system 
 of proto-neutron stars of different masses $M_1$ and $M_2$
 ($M_1+M_2=M_0$; we assume $M_2 < M_1$ hereafter). 

 These binary components begin to approach each other 
 due to the loss of total angular momentum
 and kinetic energy of orbital motion through the radiation
 of gravitational waves (GW)\cite{ImPop98}:
 \begin{equation}\label{LGW}
 L_{\mathrm{GW}}(t)\, =\,\frac{32\, G^4(M_1 +M_2) M_1^2M_2^2}
   {5\, c^5\, a^5(t)}\approx 10^{52-55}\,\mathrm{erg/s}\, ,
\end{equation}
 where $L_{\mathrm{GW}}$ is the GW luminosity, $G$ is the gravitational 
 constant, and $c$ is the speed of light.
 The mutual approach of the components lasts until the orbital radius
 $a(t)$ reaches a critical value $a = a_\mathrm{cr}$ for which
 the {\em less massive\/} component fills its Roche lobe.
 Contrary to normal stars, the degenerate configurations
 like NSs and white dwarfs have a remarkable property:
 the smaller their mass, the larger their radius ($R\sim M^{-1/3}$).
 
 As soon as $a$ becomes less than $a_\mathrm{cr}$, there begins
 a rapid mass transfer from the component $M_2$ to the component 
 $M_1$. The mass $M_2$ is rapidly decreasing down to the minimum
 possible mass of a NS $M_\mathrm{NSm}\approx \NMS{0.1}$.
 When $M_2$ becomes a little bit less than
 $M_\mathrm{NSm}$, the process of the hydrodynamic destruction of
 a low-mass component begins. Such a dynamical instability 
 is controlled by the rate of beta-processes, and initially is
 developing rather slowly. It terminates, however, with a short 
 ($\sim$0.05 s) phase of a violent
 transformation of the internal energy into kinetic energy and 
 work against gravity. The resulting energy release is expected
 to be as large as $\sim 10^{51}$erg ($\sim$4.8 MeV per nucleon). 
 However, the sophisticated calculations
 are still to be done to estimate how much of the energy 
 is taken away by neutrinos. The hydrodynamics of the low-mass 
 NS explosion was studied in a number of papers 
 (see Refs.~\refcite{Beal90,Col93} and references therein).

 The low-mass NS explosion model has no problems 
 with the explanation of the explosion asymmetry 
 (like that observed for \SNLMC) and the origin of
 the high-velocity pulsars. 
 The debris of exploded low-mass NS ($\NMS{0.1}$)
 and the collapsed NS ($\sim\NMS{1.5}$) retain their
 orbital velocities of (7500--15000) and (500--1000) km/s,
 respectively, and move in opposite directions\cite{AZIN97}. 
 There is no problem in this scenario also with
 dissociating of the heavy elements in the infalling envelope.

 Thus in this scenario, the supernova outburst is triggered
 by the explosion of a low-mass NS.
 The most impressive feature of the scenario is its ability
 to explain the two neutrino signals from \SNLMC\ in the
 framework of a single self-consistent model.
  \begin{table}[h]
 \tbl{Responses of the neutrino detectors to the $\nu_{\mathrm e}A$ 
 interaction for the Mont Blanc (LSD), KamiokaNDE II (KII), and 
  Baksan (BUST) events. 
 (Adapted from Ref.~\protect\refcite{ImRya04}).} 
 {\begin{tabular}{@{}clccl@{}} 
 \toprule Detector & Energy & Predicted & Predicted & Actually\\
 & threshold & number $N$ of & counts & detected\\ 
 & MeV & interactions & $N\eta$ \\ \colrule
LSD  & 5 -- 7  & 5.7 & 3.2     & 5\, (Refs.~\refcite{Dadyk87,Agli87})\\
KII  & 7 -- 14 & 3.1 & 2.7     & 3\, (Ref.~\refcite{DeRuj87})\\
BUST & 10      & 5.2 & $\sim$1 & 1\, (Ref.~\refcite{Alex87})\\ \botrule
\end{tabular}}
\label{tab1}
\end{table}

   The first recorded by LSD neutrino pulse comes from the 
   first stage of the collapse when there occurs the rotational
   fragmentation of stellar core into two proto-neutron stars.
   This is essentially a three-dimensional process.
   A strongly flattened by rotation structure of the core 
   favors the emission of highly non-thermal {\sl electron\/}
   neutrinos with the individual energies of (30--40)$\,$MeV
   that appear owing to capture of strongly degenerate electrons
   by atomic nuclei and free protons 
   $(p+\mathrm{e}^-\rightarrow n +\nu_\mathrm{e})$. 
   When such energetic $\nu_\mathrm{e}$'s reach the LSD detector
   they interact with 200 tons of the iron safety and radioactivity
   shield around 90 tons of scintillator (white spirit). 
   The cross-section of the reaction
   $\nu_\mathrm{e}+\isn{Fe}{56}\rightarrow\isn{Co}{56}^*+\mathrm{e}^-$
   proved to be of the order of \xmn{4}{-40}cm$^2$ -- just appropriate 
   value to interpret with a statistically good accuracy the observed 
   five LSD events as a response of scintillator to 
   gamma rays from deexcitation of $\isn{Co}{56}^*$ and to the electrons.
   Similar effect (however with a lower cross-section) occurs due to
   $\nu_\mathrm{e}$ interaction with such heavy constituents of 
   the scintillator itself  as \isn{C}{12} and \isn{O}{16}.
   Table \ref{tab1} presents the responses of different detectors
   to the $\nu_\mathrm{e}-A$ interaction ($A=$\isn{Fe}{56} for LSD 
   and BUST, and \isn{O}{16} for K~II).
   The KII and BUST detectors could not confidently detect 
   the first neutrino pulse owing to their higher than for LSD
   thresholds and backgrounds 
   (for further details see Ref.~\refcite{ImRya04}).
   
   The second neutrino pulse occurs approximately at the moment 
   of the low-mass NS explosion when more massive NS component
   $(M_1 = M_0 - M_\mathrm{NSm})$, having been already deprived 
   of a good portion of its angular momentum, resumes
   the collapse to produce a powerful burst of more or less 
   thermal neutrinos described in previous section.

   The time interval of 4.7$\,$h separating the two neutrino
   pulses is controlled by the GW radiation and can be easily 
   explained by the evolution of the NS binary system that
   is described by a semi-analytical 
   approach\cite{ImPop94,ImPop98,ImPop02}.
   
   In the future, the proposed scenario can be verified
   with the aid of the GW detectors such as VIRGO and LIGO,
   at least within the Milky Way distances of $D\approx 10\:$kpc.
   The amplitude of metric perturbation in question is 
   estimated\cite{ImPop98} to be 
   \begin{equation}\label{metric}
   h\, =\,\frac{8\, G^2M_1M_2}{\sqrt{5}\, c^4 a\, D}\approx 
   \xmn{(1-0.2)}{-18},\quad D=(10-50)\:\mathrm{kpc}\, .
   \end{equation}
   According to a conservative estimate, the gravitational
   waves carry away in total $10^{50-52}$ erg
   within a frequency range (100--3000)$\:$Hz.
   
\section{Neutrino Nucleosynthesis}

\begin{figure}
 \centerline{
\epsfig{file=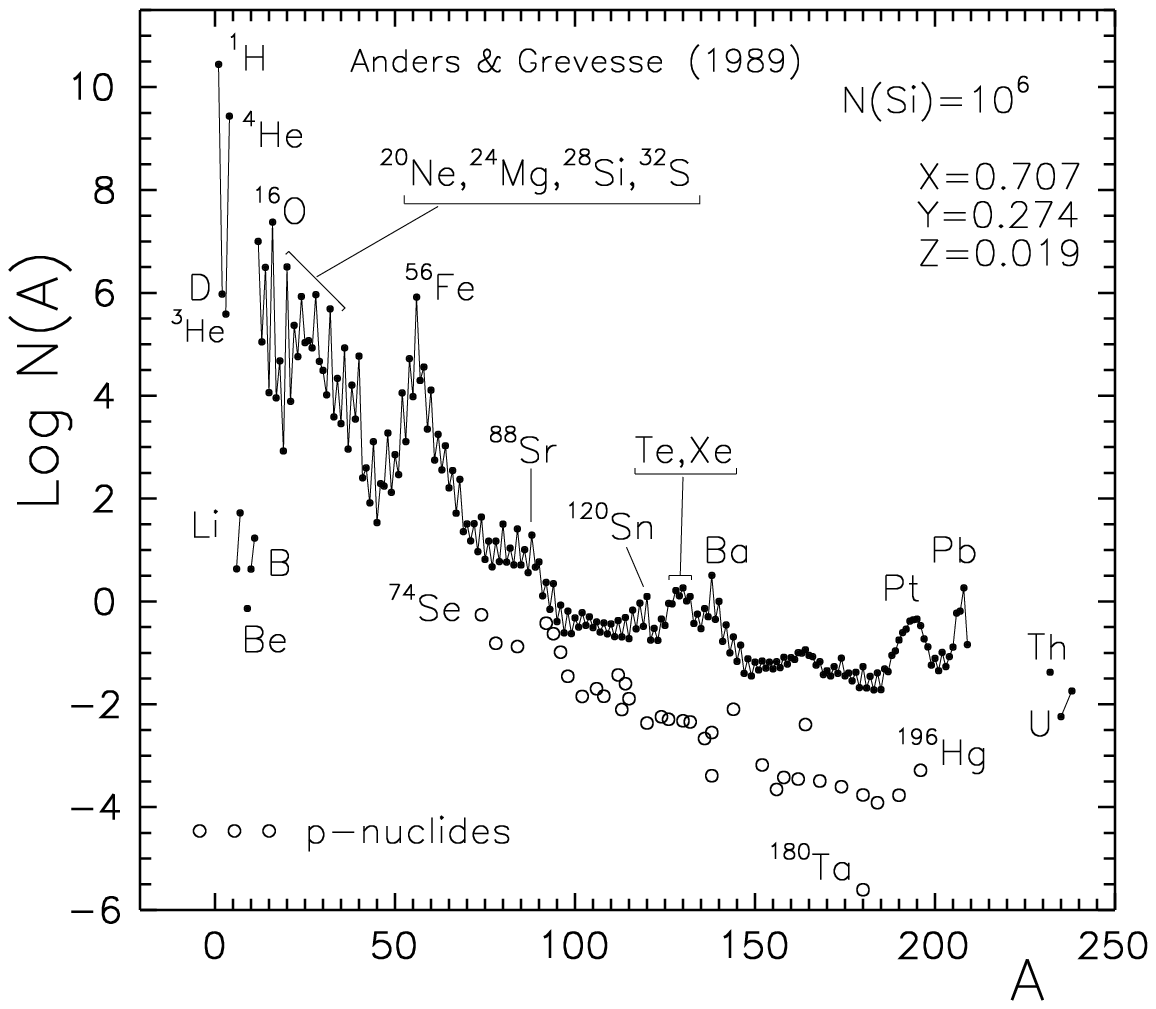,width=0.6\textwidth}\hfill
\epsfig{file=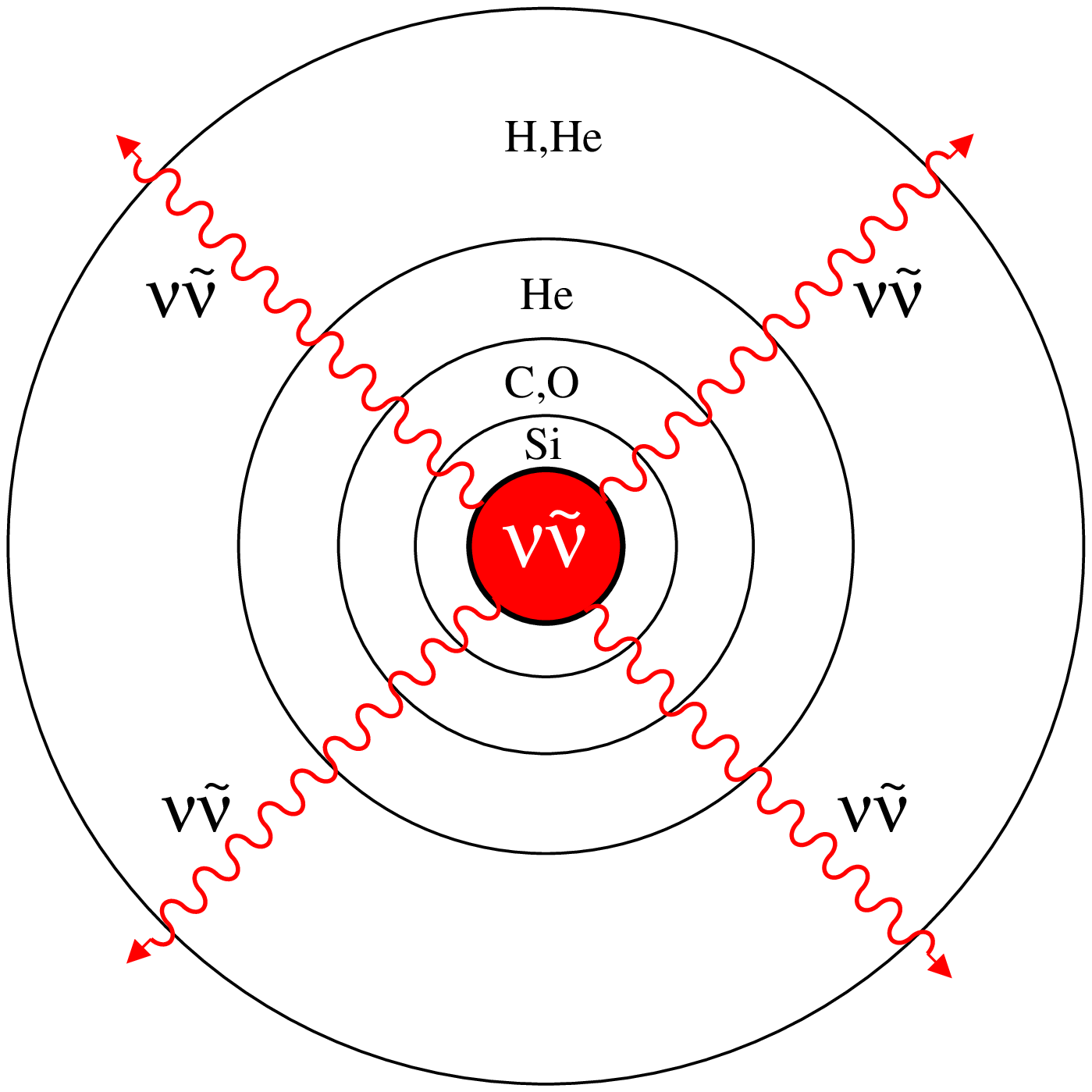,width=0.39\textwidth}
 } 
 \vspace*{8pt} 
 \caption{The cosmic abundances of chemical elements
  \protect\cite{AG89} versus the mass number $A$.
  The p-nuclides are shown explicitly (left panel).
  The onion-like presupernova structure (right panel).
  }
  \label{AGonion}
 \end{figure}
 The neutrinos from collapsed stellar cores can produce a number 
 of nuclear transmutations in the onion-like structured envelope
 (Fig.~\ref{AGonion}, right panel) to be thrown off by the blast
 wave\cite{DN78,DEN78,Woos90,Heger03}. 
 With the aid of neutrinos, it is possible to overcome difficulties
 in interpreting the cosmic abundances of such nuclear species 
 as $(i)$ p-nuclides that cannot be produced in the neutron-capture 
 processes, $(ii)$ a number of rare nuclides
 (\isn{N}{15}, \isn{F}{19}, \isn{Al}{26}, \isn{V}{50} 
 and some others), and $(iii)$ the light elements (Li, Be, B).
 The magnitude $\delta n(A,Z)$ of the neutrino-induced transformation
 in the reaction $\nu + (A,Z)\rightarrow\ldots$ can be estimated
 from a simple equation
  \begin{equation}
  \frac{\delta n(A,Z)}{n(A,Z)}=
  N_\nu\frac{\avr{\sigma_{n\nu}}}{4\pi r^2}=
  \frac{E_\nu}{\avr{\varepsilon_\nu}}
  \frac{\avr{\sigma_{n\nu}}}{4\pi r^2}\approx (0.01-0.1)\, ,
  \label{dnn}
  \end{equation}
  where $N_\nu$, \avr{\sigma_{n\nu}}, and $r$ are
  the total number of emitted neutrinos responsible for
  the transformation, their energy-averaged cross-section,
  and the distance of the neutrino-irradiated matter,
  respectively. Each of three $\nu\widetilde\nu$ flavors 
  can be involved in the transformation (\ref{dnn}).
  The numerical quantity is estimated for typical values
  $N_\nu=\xmn{3}{57}$, $r=10^9\,$cm, and
  \avr{\sigma_{n\nu}}$=$\xmn{(3-30)}{-41}cm$^2$.
  
  The creation of the light elements is a field of
  overlap between the contributions from cosmic rays (CR),
  big bang nucleosynthesis (BBN) 
  and neutrino nucleosynthesis (NN). The neutrino flux
  proves to be especially efficient in production of 
  \isn{Li}{7} (in helium shell) and \isn{B}{11}, \isn{Be}{9}
  (in CO-shell). At the same time, it is not so effective in
  producing of \isn{Li}{6} and \isn{B}{10}. 
  Thus, the large cosmic ratios\cite{AG89}
  \isn{Li}{7}/\isn{Li}{6}$=12.3$ and 
  \isn{B}{11}/\isn{B}{10}$=4.02$ can be easily understood 
  in the framework of neutrino nucleosynthesis.
  And there is no need to invoke a hypothetical low-energy
  ($E\lsim 100\,$MeV) cosmic rays to explain these ratios.
  Presumably, cosmic abundances of the light element isotopes
  can be considered as coming from combinations of contributions
  from BBN, CR, and NN:  \isn{Li}{6} (CR), 
  \isn{Li}{7} (BBN$+$CR$+$NN), \isn{Be}{9} (CR$+$NN), 
  \isn{B}{10} (CR), and \isn{B}{11} (CR$+$NN).
  One has also to keep in mind other possible stellar sources 
  of the light elements, e.g. such as novae and red giant (AGB)
  stars.
  
\section{Gamma-Ray Burst -- Supernova Connection}

 Cosmic source of the gamma-ray bursts (GRBs) discovered 
 30 years ago\cite{Kleb73,Maz74} long remained actually 
 unknown. The breakthrough in understanding of the GRBs 
 happened in late 90ths. There was discovered\cite{Costa77}
 that at least some GRBs were followed by afterglows.
 Then the afterglow was found to be connected to 
 supernovae\cite{Kulka98,Galama98}. This findings allowed
 to couple GRBs with host galaxies of known redshifts and
 thereby directly to confirm previously assumed cosmological
 distances (of hundreds Megaparsecs) to the GRBs.   
 The most convincing proof of the GRBs--SNe connection 
 came from Chandra's 21-hour (!) X-ray exposure of 
 the afterglow associated with the GRB detected on 
 13 August 2002 that allowed to identify narrow lines
 due to silicon and sulfur ions inherent 
 to the SN ejecta\cite{Butler03}. 
   
   The GRB mechanism is not yet well understood.
   However, the gamma-rays are thought to be beamed into
   a narrow cone along a jet of high energy relativistic
   particles expelled from a supernova core that has just
   collapsed into either a NS or a black hole
   (see Refs.~\refcite{Woos93,Blin00,Post04,WH2004} and 
   references therein). Such jets seem to occur 10--60$\,$d 
   after the beginning of the collapse. 
   They interact with the expanding supernova envelope 
   and produce the X-ray and optical afterglows\cite{Matz03,Kosen03}.
   The core-collapse SNe Ib and Ic are expected to be the best
   sites for creating the GRBs. These SNe are deprived of dense
   hydrogen-rich envelopes that would prevent gamma-rays to
   escape from the star.
   
   The jet-like streams of relativistic particles can be 
   a good starting point for writing a new chapter in the
   theory of origin of cosmic rays\cite{Dar04}.
   The GRBs--SNe connection opens a new intriguing approach
   to understanding the mechanism of the core-collapse SNe.

\section*{Acknowledgments}

 D.K.N. has a pleasure to thank the Institute for
 Nuclear Research of Russian Academy of Sciences and
 Laboratori Nazionali del Gran Sasso of Istituto Nazionale
 di Fisica Nucleare (Italy) and their staffs
 for warm hospitality during his staying in Italy. 
 The work was supported by the Russian 
 Foundation for Basic Research (project no. 04-02-16793-a).

\end{document}